\documentclass{kluwer}
\begin{document}
\begin{article}
\begin{opening}

\title{In ISM Modeling, The Devil is in the Details: You Show Me Your OVI and I'll Show
You Mine.}
\author{Donald P. \surname{Cox}}
\runningauthor{Don Cox} \runningtitle{The Devil is in the Details}
\institute{Department of Physics, University of Wisconsin-Madison,
1150 University Ave., Madison, WI 53706, USA, Email:
\texttt{cox@wisp.physics.wisc.edu}}

\begin{abstract}
The three sections of this paper
illustrate the importance of rationalizing ISM
theory and modeling with the observational information.

\end{abstract}
\end{opening}

\section{Modeling the ISM is Complicated.}

On this topic, see also Maarit Korpi's contribution  and 
two other papers of mine (``Guide to Modeling the
Interstellar Medium,'' and ``Overall Models of the Interstellar
Medium,'' 1996a,b).

Imagine pouring the ISM
into the Galaxy's gravitational potential, where it finds all
the other constituents already in place.  It is
globally confined by gravity, stabilized radially by rotation,
heated by starlight, disrupted by catastrophic stellar events,
settles into components spanning a wide range of density and
temperature, develops a magnetic field, confines cosmic rays,
generates new stars, interacts with stellar density waves, and
is concentrated by that into the gaseous spiral
arms.  It develops a vertical quasi-hydrostatics, with dynamic
pressure, magnetic pressure (and tension), and cosmic ray
pressure as dominant forms of support.  When locally
heated, it develops bubbles of hot gas that grow until their
thermal pressures are in approximate equilibrium with the
magnetic pressure that surrounds them.  This can occur either
via random heating events (accidental sequences of supernovae
in proximity), or via the local production of OB associations
that act collectively to
build superbubbles.

Modeling this situation is approached at a variety
of levels, e.g. Processes, Events, Local Region, and Global.
At intermediate levels, one assumes that the processes at lower levels
can be included in subgrid rules, while those 
at higher levels are  embodied in initial or boundary conditions.  
One models the remnants of supernova explosions, for example,
assuming an adequate understanding of 
the atomic and radiation processes, magnetic fields, thermal conduction,
dust destruction and cosmic ray acceleration by shocks, etc.,
and a sufficient description of the surrounding medium.  The results
and their regime of applicability depend on the way in which all these
lower and higher level aspects are approximated.  The validity
tends to increase with the degree to which the models are compared 
in intricate detail with the suite of observational data.  
With care, some parts may turn out to be right.

The assumptions we make are often those of convention;
occasionally this may lead us astray. 
In general, astrophysicists are not comfortable with thermal 
conduction, for example.  Magnetic fields interfere with
it.  Turbulence alters its basic nature (Lazarian and Cho,
this volume).  It is easy to leave it out of models because
so many people believe it is quenched, while the others 
offer no clear prescription for how to include it.  
And yet it is a dangerous thing to omit in
models that include hot gas.  In two systems that I helped
model, the Local Bubble and the W44 supernova remnant, the
apparently observed characteristics appeared
only when thermal conduction was included within the hot gas 
at roughly half the Spitzer rate.  I've heard that
something similar is being found in modeling cooling
flows in clusters.

It is not clear how robust this hint is, but it begs
for consideration.  If you model the ISM, and 
find large volumes of hot gas, then please,
to prevent our wasting too many more years in ignorance, add
thermal conduction (with a variable amplitude parameter and
limited by saturation) and see what its effects are.  
If your resolution is high enough that the 
turbulent and wandering field is adequately represented,
let conduction cross your magnetic fields, with a reduced coefficient,
perhaps of order 5\% of the parallel rate (or ask Randy Jokipii what
he would do).  My
prejudice is that thermal conduction in high temperature gas
somehow manages to occur, despite magnetic fields, but that it
is unable to cross boundaries to denser regions
and thermally evaporate them.  Maybe your models will give us
a better sense of why.

Modeling superbubbles has a
long history, including an early fascination with
bubbles breaking out of the gaseous disk to dump their hot
interiors into the Galactic halo.  When it was
finally realized that the gaseous disk is very thick in low density
and nonthermal components, the fascination went away.  
Even as large as they are, the bubbles are confined within the 
the disk, which they help inflate,
consistent with x-ray observations of other galaxies that find
very few plumes of hot gas rising out of their disks.  The 
fascination with breakout was predicated on misinformation at 
the larger scales.

But modeling superbubbles may have problems beyond finding
a reasonable representation of the ambient medium.   
Have we got the process details of the interior
right?  Do the winds of the OB stars, augmented
perhaps by entrained material from the molecular cloud
birthplace, hit the shell and set up a reverse shock,
applying the accumulated pressure to the bubble walls?  
Or is the material so dilute that there is no collisionless
shock in the wind?  Does the fluid description give us the
wrong answer?  (I ask because Sally Oey is
trying to understand why so many wind bubbles appear smaller
than they theoretically ought to be.  See her contribution to this volume.)  
We have great theories, but they do not encompass the
observed facts.  (But, Don Osterbrock would often say something like, 
``Never let questionable observations interfere
with your appreciation of a good theory.''  The corollary that
most observations can be questioned
is one of Sally's approaches.)

In short, there are areas of ISM research in which important
details are too poorly known for us to be confident
in the ways we include them in models at other
levels.  That said, I am enormously impressed by the
incredibly detailed models that are starting to appear.  This
meeting was very good in presenting the remarkable ambition of
the current enterprises.

\section{There Are Many Conceptions of the ISM, All Flawed.}

What pictures would you draw to illustrate
what you think are the dominant features of the ISM?  I ask because 
what you draw determines what you put in your computer model,
or how you describe the significance of your observations.

There are several commonly held
conceptions, at different scales.  Let's begin with
conceptions within the disk, near us.  The ones I
will describe are the trivial ones.  People such
as Miguel de Avillez, who make models of the medium based on
their best guesses of the process behaviors might draw
something much more complex.  One might even appear on the
cover of this book, as it did on the book of abstracts.

Local conceptions tend to diverge over the component of the
medium occupying the greatest volume.  The three favorite
possibilities appear to be:

\begin{itemize}

\item A warm intercloud medium, density perhaps averaging about
$0.2$ cm$^{-3}$ and temperature perhaps $7000$ K, in some places
ionized and others not.

\item A pervasive hot component with a temperature of roughly
$10^6$ K and pressure $p/k \sim 10^4$ cm$^{-3}$ K.

\item A tepid intercloud component with a temperature of $1-3
\times 10^5$ K.

\end{itemize}

The first of these views derives from the legacy of Field,
Goldsmith and Habing (1969).  It is somewhat supported
observationally by the fact that there are warm HI and diffuse
HII components that might be able to fill most of the volume.  
McKee and Ostriker (1977) argued that supernovae would
disrupt such a medium and that the picture was therefore
untenable.  Slavin and Cox (1993), however, located the three
fallacies leading to that conclusion, showed that supernovae
might not disrupt such a medium, and found that invisible old
remnants within it could provide all of the observed high
stage ions (e.g. OVI) found in absorption measurements through
the disk.  Spitzer (private communication) however, wondered
how the warm intercloud medium might go about restoring its
relative uniformity.  That question aside, it
remains a somewhat attractive possibility, even though the
detailed reasons for why it might be about right are unclear.

The second possibility follows from a suggestion by Cox and
Smith (1974), and one much earlier by
Spitzer (1956).  The soft x-ray background appearing to derive from a large
region of million Kelvin gas surrounding the Solar location
led Cox and Smith to ask whether such gas might 
be common.   I have abandoned
this possibility because I think there is no
observational evidence to support it, as well as no clear
reason why it might be the preferred situation.  I could be
wrong; my distaste hinges
largely on the expectation that it would have shown up more
convincingly in soft x-rays.  If, on the other
hand, it were a few times hotter in most places, at a
correspondingly lower density (to not have an unacceptably
large pressure), its x-ray emission would be negligible and
the scenario observationally allowed.  (I like to refer to
this as the Trans-Mega-Kelvin Regime.)

McKee and Ostriker (1977) suggested that 
there would likely be a thermal runaway in the hot
component of Cox and Smith, unless one invoked a 
thermostatic mechanism such as
evaporation of clouds to add mass.  
In this way, they evolved to the third scenario
above.  Their medium is heated to very high temperatures 
in local regions by supernovae, evaporates clouds,
lowers its temperature as it expands and gains mass, and
finally is able to radiate the supernova power when the
temperature has dropped to a few hundred thousand Kelvins.  
The observed OVI was said to be
in the thermally evaporating boundaries of the clouds, with
the right average column density.  The fact that far more OVI
would be present in the tepid ambient medium, more than is
observed by a large factor, was overlooked.  For
years I was convinced that the MO model was
inconsistent with observation, but could not figure out how it
failed.  As remarked earlier,
Jon Slavin and I finally found sufficient reason.  It was a
great example of observations insisting that the theory was
lacking.

There is a fourth possibility that I try to remember to
mention, thanks to a casual remark by
Priscilla Frisch, in 1986.  Perhaps most of the space is
empty!  This appears to be the impression of observers
who study galactic sightlines and find components that  fill only 20 or
30\% of the total line of sight.  Their copout, if they
mention it at all, is that the rest of the sightline must be
occupied by hot gas as advocated by . . . someone
else.

I'll try to make empty space sound not so crazy.  It isn't
that there is nothing in the empty regions.  The
magnetic field and cosmic rays are still there; 
the field lines just have very little mass on them.  It's like
an extreme version of the Trans-Mega-Kelvin
regime.
There is no observational reason to reject
this conception; we just have no theoretical
framework in which it is plausible.  How would such empty
regions come into existence (McKee and
Ostriker's thermal runaway perhaps)?  
How do they evolve?  Do they exit the
Galaxy because of their buoyancy, or evolve out of existence
via material leaking (or being driven) into them?
Sound crazy still?  Well, the last I heard, the outer shock of the Crab
Nebula had not been found.  The explosion appears to be
expanding into a vacuum.

With regard to the more complicated pictures being drawn by
large scale modelers, I can only agree with Joel Bregman that until
they are able to tell us how much OVI is in their descriptions, we
can't look at them with a sufficiently appreciative eye.  
I'll wager they get at least an order of magnitude too much!  It
is a powerful applicability check, hence the subtitle of this paper.

Then there are the vertical and the global conceptions.
If the galaxy were truly filled with hot gas 
(not tied to a magnetic field anchored by the weight of the cold gas),
such gas would extend far above the disk of the Galaxy,
filling out a scale height at least.  
If many superbubbles grow so large that they
break out of the disk, the hot gas from their interiors
might plausibly fill the halo, the density rising until the
hot gas is able to radiate the supernova power.  Or, if the
escaping gas were fairly dense, it might
generate a fountain, cooling before it gets very far,
condensing into clouds that eventually rain back.  
So, we have two scenarios, a hot quiescent halo, and
a fountain.  But we know from observations that the halos of
galaxies do not radiate even a small fraction of the total
supernova power, not at high temperatures.  Both scenarios
failed badly, to the great surprise of many of us who were
expecting otherwise when x-ray data first became available.
  
What's left then?  Well, cosmic rays escape the Galaxy.  
The phenomenon is frequently
described as a wind.  Perhaps the Galaxy has a
thermal wind too, arising from the disk, high enough in
temperature (and therefore low enough in density) that its
x-ray emission is negligible (the Trans-Mega-Kelvin regime again).
I believe that such a wind cannot be carrying away thermal energy
faster than the cosmic ray loss rate, which is roughly 10\% of the
total supernova power.
The cosmic rays seem to diffuse out of the disk, 
forming their ``wind'' rather high off the plane. 
If hot gas flows out rather unimpededly, it ought to
take cosmic rays with it, in which case the two would leave the
Galaxy in proportion to their relative energy densities in the 
hot component in the disk, a ratio of order unity in this scenario.

An entirely different possibility is that except for 
cosmic rays leaving the Galaxy, all other
disturbances are trapped in the disk by the inelasticity of
the great interstellar feather pillow.  Superbubbles are the
largest disturbances, but few reach such great
size they are other than trapped volumes of hot gas that
eventually dissipate their energies radiatively.  
This has been my favorite view
for some time.  It does not mean that there is no hot gas above
the plane of the Galaxy; there are superbubbles that expand
several hundred parsecs.  We see several, and
hot gas within them.  There are also supernovae that occur at high
$|z|$.  These may be an important contributor to the thick
layer in which high stage ions are found, as explored by Robin
Shelton (1998).  It just means that violent disturbances are
generally locally confined, and dissipated.

What picture would you draw?  Mine has all the usual
components.  
There is a thick layer of low density warm
intercloud gas, reaching up a kiloparsec or two, with here and
there a confined supernova remnant within it.  Quite a bit of it
is ionized and visible in H alpha, the Reynolds Layer.  It is
pervaded by a largely random magnetic field and cosmic rays.  
Within it, dense cold clouds lie close to the midplane creating
OB associations, the latter forming superbubbles that expand,
but not to the extent that they break out of the disk.  To be more
precise, I think they help inflate the disk, rather than breaking out of it,
except in the extreme case of starburst galaxies.
Then there is the emptiness, emptiness that fills the ancient
and dying carcasses of once grand superbubbles, \`{a} la the
models of Ferri\`{e}re (1998), and maybe even the emptiness of
Frisch or the Trans-Mega-Kelvin regime.  And finally, the
whole thing is circulating around the Galaxy, now and then
provoked into unusual activity by its encounter with a spiral
arm.
This spiral arm encounter is often described as a shock,
but, Martos and Cox (1998) 
showed that it should have some elements of
both a shock and a hydraulic jump, or bore.  
That work has been extended to three dimensions
and a significant fraction of the Galaxy by G\'{o}mez and Cox
(2002).  A brief report appears in the poster contributions to
this volume.  It's exciting and an excellent
example of a large-scale model that has to make very simple
approximations to the character of the medium.
There could also be a wind from the
Galactic Bulge, providing a very interesting
outer boundary condition on the ISM of the disk.

\section{Local Matters Matter.}

This is about building a house of cards.  It
concerns the structure of the ISM in the vicinity of the Sun,
starting out about 100 pc away and working in to about 5 AU.  
It is a progress report on years of work, and the fearful
tests that lie ahead.  Parts of it are discussed at length in
Smith and Cox (2001) and Cox and Helenius (2003).    
Other authors have taken different routes with the
observational data; references and discussions of their
views appear in the papers above.  But I am here
offering an example of a modeling effort that has gone out on
a limb, not surveying the whole realm of possibilities.


The Local Cavity is a large region of exceptionally low density surrounding
the Solar location.  
Within it, there appears to be a sub-region
of million Kelvin gas, the Hot Local Bubble, providing much of the soft x-ray
background.  To get the observed
x-ray surface brightness, given the scale of order 100 pc, the required
thermal pressure is $p/k \sim 2 \times 10^4$ cm$^{-3}$ K.
This is just
what would be expected for such a bubble quasi-statically
confined by the surrounding ISM, where the pressure of the latter equals its
weight per unit area.
Within that hot gas, there are wisps of higher density
cooler material, the Local Fluff.  The densities in the Local
Fluff are of order $0.2$ cm$^{-3}$, the temperatures about 7000 K,
characteristic sizes of about 5 pc.  One particular piece of
this Fluff, the Local Cloud, surrounds the Solar System.  
The relative velocity between the Local Cloud
and the Sun is about $26$ km s$^{-1}$.  Thus, in the frame of the
Sun, the material of the Local Cloud (the LISM, and VLISM),
constitutes an interstellar wind blowing past the Solar
System.

The Solar wind interacts with this interstellar wind, 
its momentum creating a large
cavity around which the flow of interstellar ions and magnetic field
is diverted.  To first order, the neutral component of the
Local Cloud does not notice this disturbance.  The scale is
sub-fluid for the neutrals, which tend to flow undisturbed
into the heliosphere.  (There is some charge exchange between
the neutrals and diverted interstellar ions that has important
consequences for the details.)  Once they are deep into the
Solar System, these neutrals suffer various fates.  When one
becomes ionized, it is immediately picked up by the
Solar Wind magnetic field and joins the outflow; these are
observed as ``pickup ions'' by interplanetary probes.  On reaching
the Solar Wind termination shock, some are accelerated to
higher energies, diffuse back in toward the Earth, and are
observed as the ``anomalous cosmic rays.''
Radiation pressure from
scattered Solar Lyman alpha approximately cancels
gravity, so the hydrogen trajectories are straight.  
Solar radiation ionizes most of it
before it reaches 5 AU.  The distribution
of interstellar hydrogen in the inner Solar System is
therefore roughly uniform in the upstream direction, outward
of 5 AU, and absent at smaller radii and in much of the downstream
direction.  It is observed via its backscattering of the Solar
Lyman alpha.
Helium suffers less radiation pressure and ionization,
and is gravitationally focused in the downstream
direction.  This is observed via
resonant backscattering of 304 {\AA} Solar radiation.


Smith and Cox (2001) explored the possibility that the Local
Bubble could have been created by an accidental sequence of
supernovae in a warm intercloud medium as described in Section
2.  They found that a sequence of three supernovae over the
course of several million years, the second and third each
occurring within the region previously evacuated, could create
a bubble of radius about 100 pc, and leave it with the right
density and temperature so that its radiation would resemble
the soft x-ray background.  Considering that the Local Cavity
is much larger in some directions, and potentially much older,  
a generalization is that something
created the large cavity, and a supernova reheated a portion
of it roughly three million years ago to leave us surrounded
by the Hot Local Bubble.  Evidence for the nearby
occurrence of a supernova about that long ago is discussed in
their paper and further in Cox and Helenius(2003).

But why would the Local Fluff be sitting in the middle of such a remnant?  
A possible answer was provided by Cox and Helenius (2003) who examined the
evolution of a magnetic flux tube pulled from the wall of the
bubble by its magnetic tension, starting about three million
years ago, after it had been swept to the wall by the supernova.
As the flux tube springs from the wall, it is initially
accelerated very rapidly; but soon drag through the hot gas
becomes important, and it gradually decelerates.  
Material within the tube starts out very diffuse, part of
the inner wall of the bubble, but the initial acceleration due
to tension is largely radial.  As the tube itself slows,
the material within it flows toward
its center, concentrating into a low-density cloud, or
collection of interacting clouds, depending on the complexity
of the initial conditions.  By the time the tube reaches the
bubble middle, the velocity distribution of the material
within resembles that observed in the Local Fluff, in
magnitude, and in the generally converging flow pattern.  Its
density and ionization state are also appropriate.   The details 
need more study, but the scenario
creates something very much like the observed
Local Fluff.


Now we turn to the problems.
Understanding of the soft x-ray background is not
complete.  High spectral
resolution will eventually give us a much better idea of what we are
looking at.  For the present, there is at least one
disturbing possibility.
The sky maps of the SXRB from ROSAT were made from data that
had contamination that was not
understood at the time, the ``Long Term
Enhancements.''  In mapping, periods of time during
which this contamination was large were omitted.  The maps
thus present the lowest signal detected from each
direction.  Their spatial
structure correlates very well with maps from rockets,
and Galactic structure, so it is believed that the process was
successful in removing the contamination.  
But the resulting 1/4th keV  surface brightness is pretty uniform
in the Galactic Plane, getting brighter at higher latitudes.  
The worry is that the 
midplane intensity could be just the DC level of the
contamination.  If so, we would lose the Local
Bubble as we think we know it, replacing it with smaller
bubbles at positive and negative higher latitudes.
After the observations of x-rays from
comets, and their interpretation as arising from charge
exchange recombination of highly ionized elements in the Solar
Wind, we finally had a mechanism for generating x-rays within
the Solar System.  It now seems likely that  the Long Term
Enhancements arise from the same mechanism, with the
charge exchange occurring on the atoms of the Earth's
exosphere.  There should be also be a
significant steady contribution from the Solar Wind ions charge
exchanging on the interstellar neutrals penetrating the Solar
System (e.g. Cox, 1998, Cravens et al., 2001).  
The optimistic view from the estimates so far is that the level of the overall
contamination is modest, at  30\% or so, and that our
picture will remain largely undisturbed.  We'll
see.

The other major problem is the distance to the Solar Wind
Termination Shock.  I told you what the pressure in the Local
Bubble had to be to produce the observed x-rays, and that it
is appropriate for a bubble in equilibrium
with the surrounding ISM.  I also gave you data from which you
can estimate the thermal pressure in the Local Fluff.  
The latter is less than 20\% of the
former.  In the model of Cox and Helenius, the magnetic pressure
of the flux tube is significantly higher than the
thermal pressure of the gas within it.  The field strength is
several microGauss, to balance the thermal pressure of the
surrounding hot bubble.  Forget the model.  In order for the
Local Fluff to be in pressure equilibrium with the
surroundings, such a field is required.
But modelers of the interaction of the Solar Wind with the
Local Cloud will tell you that if the magnetic
pressure in the Local Cloud were that high, the termination shock
would have been pushed inward to the point that Voyager 1
would already have encountered it, which it has not!
There are only a few ways out of this difficulty.  One is that
making such models is difficult (as in the theme of Section 1)
and details may have been overlooked.  Astronomically 
insignificant things like a 
measly factor of two are critical.  Another
is that the Cox and Helenius model makes a specific prediction
for the direction of the magnetic field, from the
directionality of the local distribution of Fluff, and that
direction is roughly parallel to the relative wind flow.  This is a
case that has not been examined and offers a radically
different structure.

A third alternative is the scary one, involving both of the
above problems.  The field is actually much weaker.  The Local
Bubble does not exist.  The local interstellar total pressure
is very significantly lower than the average expected from the
weight of the interstellar medium.  The whole house of cards
collapses.  We'll see.

\vspace{0.5cm}

One last thing:  If you model the ISM,  don't forget to tell us how much OVI you get.
Send me a telegram when you no longer find more than a factor of two too much!

\vspace{0.5cm}

This work was supported in part by NASA grant NAG5-8417 to the
University of Wisconsin-Madison.  It has benefited from critical readings by 
Gilberto G{\'o}mez, Bob Benjamin, Sally Oey, Maarit Korpi, and Dieter Breitschwerdt.  
I apologize to the many
observers whose work made the above discussion possible yet
who were not mentioned by name or referenced.  To others,
please do not suppose that I have told you the truth about observations. 
Go to the sources.

\end{article}
\end{document}